# Superconductivity in Undoped Single Crystals of $BaFe_2As_2$: Field and Current Dependence

## J. S. Kim, T. D. Blasius[†], E. G. Kim and G. R. Stewart[*]
### Department of Physics, University of Florida, Gainesville, FL 32611-8440

Email: stewart@phys.ufl.edu; Phone: 352 392-9263; Fax: 352 392-8586

**Abstract:** In previous work in undoped $MFe_2As_2$, partial drops in the resistivity indicative of traces of superconductivity have been observed in some samples of M=Ba ($T_c \sim 20$ K, up to 25% drop in $\rho$) and M=Ca ($T_c \sim 10$ K, up to 45 % drop in $\rho$.) A complete drop in the resistivity to $\rho=0$, along with a finite fraction of Meissner flux expulsion, has been observed in M=Sr, $T_c = 22$ K. Using In-flux grown single crystal samples of undoped $BaFe_2As_2$, we find a complete drop in the resistivity to 0 in most samples beginning at $T_c^{onset} = 22.5$ K. However – in contrast to the $SrFe_2As_2$ results – there is no measurable Meissner effect and no suppression of the resistive superconducting transition with annealing. The current sensitivity of the superconducting resistive transition in our samples of $BaFe_2As_2$ is quite strong, with an increase in the current density of a factor of 15 to $\sim 1.5$ A/cm$^2$ not changing $T_c^{onset}$ but broadening the transition significantly and causing $\rho$ to remain finite as $T \to 0$. To investigate if this unusually low critical current is indicative of filamentary conduction lacking the apparent anisotropy seen in the critical magnetic field, $H_{c2}$, measurements of, e. g., the bulk superconductor Co-doped $BaFe_2As_2$, $H_{c2}$ was measured in both crystalline directions. These $BaFe_2As_2$ samples show $H_{c2}(T)$ values in the ab-plane and along the c-axis comparable to those seen in $BaFe_{2-x}Co_xAs_2$, which has a similar $T_c$. Since the lack of $T_c$ suppression after annealing argues against strain-induced superconductivity as proposed for the other undoped $MFe_2As_2$ materials, another possible cause for the superconductivity in $BaFe_2As_2$ is discussed.

PACS numbers: 74.70.-b, 74.70.Dd, 74.25.Fy, 74.25.Sv

## 1. Introduction

The discovery of superconductivity in the iron pnictides has caused significant interest [1] in the scientific community. After a $T_c$ of 55 K was achieved [2] in F-doped SmFeAsO, in the so-called "1111" iron arsenic structure, superconductivity was found [3] in a new class of compounds (the "122" structure) at 38 K in K-doped $BaFe_2As_2$. Many dopants on the Ba-site other than K have since been found to suppress the spin density wave transition in the 122 parent compound, $MFe_2As_2$ (M=Ba, Sr, Ca, Eu), and cause superconductivity. In addition, doping on the Fe site with, e. g. Co [4], as well as doping on the As site with, e. g. P [5], have been found to achieve the same result.

One of the more intriguing results to date in the 122 iron pnictides is the occurrence of *partial* superconducting transitions in the undoped parent compounds: $BaFe_2As_2$, where in one work [6] $\rho$ in two out of five samples falls up to 25 % starting at ~ 20 K; $CaFe_2As_2$ where $\rho$ in at least one sample has been seen [7] to fall by almost ½ although at the much lower temperature of 10 K; and $SrFe_2As_2$, where the resistivity, $\rho$, actually goes to 0 along with partial diamagnetic screening [8] at $T_c$ ~ 22 K. The explanation to date of this behavior [6-8] has been lattice distortion/strain, i. e. a sort of an effective pressure-induced superconductivity in a small fraction of the sample. Annealing of the superconducting samples of $SrFe_2As_2$ at 200 $^o$C for 5 minutes was found to decrease the drop in $\rho$ below an unaltered $T_c$ by ~50 %, while annealing at 300 $^o$C for two hours destroyed all traces of the superconductivity [8].

We report here on characterization of single crystals of $BaFe_2As_2$ grown in In-flux [9-10] with residual resistivity ratios of between 3.7 and 5.0, which are only slightly higher than values around 3.5 in self-flux grown crystals reported [6] previously. The majority of these In-flux grown crystals show a full drop in their resistivity, with significant sample dependence in both $T_c^{onset}$ (19-23 K) and the temperature where $\rho \to 0$ (7-19 K).

## 2. Experimental

Since Saha et al. [8] find that annealing their $SrFe_2As_2$ crystals at 200 $^o$C for 5 minutes markedly degrades the superconducting transition, it is important to note the different thermal history in their growth of $SrFe_2As_2$ crystals vs that for our In-flux grown $BaFe_2As_2$ crystals. Growing [9-10] in In flux involves a slow cool from 1000 $^o$C down to 500 $^o$C, followed by a 75 $^o$C/hr cool down to room temperature. As well, removing the sample from the In flux involves heating on a hot plate to ~ 200 $^o$C for 5 to10 minutes, followed by curing of Epo-tek H31LV Ag-epoxy resistivity contacts at 120 $^o$C for 40 minutes. The thermal history of the self-flux-grown $SrFe_2As_2$ crystals involves growth [8], [11] in an FeAs flux by cooling from 1100 to 900 $^o$C at 4 $^o$C/hr followed by cooling at ~ 250 $^o$C/hr (furnace shut-off) down to ~ 400 $^o$C and approximately 50 $^o$C/hr thereafter.

A second issue to emphasize here is that the crystals we have obtained from this first growth batch (i. e. not under optimized conditions) of $BaFe_2As_2$ from In-flux are quite small, typically 1 mm on a side and 0.1 mm thick. Thus contacting these crystals was done under a microscope and the geometrical factor necessary to determine absolute resistivity values is only accurate to about 25%. The residual resistivity ratio, RRR (=$\rho$(300 K)/$\rho$(T→0)), is however quite accurate since the geometrical factor cancels in the ratio.

Finally, as also reported by all the other works on such superconducting 'indications' in the $MFe_2As_2$ superconductors [6-8], there is a certain uncontrolled sample dependence present in these results which may be linked to the as-yet poorly understood cause of this superconductivity. For example, when changing contacts on the surface of one of our crystals some material on the surface was accidentally stripped away due to their micaceous nature. The sample afterwards showed a *narrower* transition, with the temperature where $\rho$→0 increased by several degrees. Thus, either the surface is important or the reheating to 120 $^o$C when reapplying new epoxy resistivity contacts caused this change.

Resistivity was measured using a four contact dc method, with the current switched in direction for a total of 40 measurements in each current direction at each temperature. The current is supplied by a Keithley 220 current source and the voltage is measured by a high-sensitivity, low noise Keithley 2001 voltmeter. Critical field data were taken up to 8 T with the field both in and perpendicular to the ab-plane. Determining $T_c$ as either the midpoint or the onset of the resistive transition did not change the value of the slope of $H_{c2}$ at $T_c$.

## 3. Results and Discussion

The resistivities at low temperatures of five samples of In-flux grown single crystals, current in the ab-plane, of $BaFe_2As_2$ are shown in Fig. 1. Although three of the samples show complete resistive transitions to $\rho=0$, none of the samples show any dc magnetic susceptibility indication of superconductivity at the resistive transitions, in contrast to the results [8] for $SrFe_2As_2$.

Clearly, there is a wide range of normal $\rho_{ab}$ extrapolated from above the superconducting transition (~0.07 to 0.64 m$\Omega$-cm) which, at least for the superconducting samples we have measured and within the ±25% geometrical uncertainty mentioned above, appears to be correlated with $T_c$ onset: the smaller the normal state resistivity, the higher is $T_c$. However, our result for sample #5 spoils this tentative correlation, since it is not superconducting. Also, the literature values for normal $\rho_{ab}$ (T→0) in self-flux grown single crystals of $BaFe_2As_2$ are certainly comparable to the values reported here, e. g. [6] report values between about 0.06 and 0.1 m$\Omega$-cm and the samples with traces of a superconducting transition have the *larger* values, while [12], [13], [14] (all with no trace of superconductivity) report $\rho_{ab}$ (T→0)~0.15 m$\Omega$-cm, 0.4 m$\Omega$-cm, 0.6 m$\Omega$-cm respectively. Thus, there does not seem to be a basis for associating the occurrence of superconductivity with the values of the normal $\rho_{ab}$ (T→0). This is consistent with arguments for the nature and cause of the superconductivity presented below.

Fig. 2 shows the sensitivity of the superconductivity to current: 1.5 mA through the cross section of sample 1 corresponds to a current density of only 1.5 A/cm$^2$. This rather small value having such a large effect on the superconductivity caused us to consider whether the superconductivity in these samples might be filamentary, with perhaps the filaments lacking the apparent anisotropy near $T_c$ of the critical magnetic field seen in bulk [15-16] and film [17] samples.

One way to check this is to measure the critical field behavior of the resistive transitions; such data for one sample are shown in Fig. 3. Clearly, the superconductivity in our In-flux grown crystals of undoped $BaFe_2As_2$ possesses the same apparent anisotropy $\gamma$ ($\gamma=H_{c2}^{ab}/H_{c2}^{c}$) near $T_c$ as bulk [15] Co-doped $BaFe_2As_2$. In fact, not just the $\gamma$ ratio but also the values themselves of $H_{c2}$ in the ab-plane and in the c-axis direction for our undoped $BaFe_2As_2$ are comparable both to those [15] of the bulk superconductor and to those determined for the partial superconductivity seen [8] by Saha, et al. in undoped $SrFe_2As_2$. Whether or not this apparent anisotropy ratio is indeed due to real crystalline anisotropy or is rather due [17] to multiple bands with different anisotropies, the data in Fig. 3 argue against filamentary superconductivity.

What then is the origin of the superconductivity in In-flux grown crystals of $BaFe_2As_2$? It does not appear to be strain related, as postulated for the other undoped $MFe_2As_2$ 'partial' superconductors, since the same annealing regimen (300 °C for 2 hours) that [8] totally suppressed the superconductivity in $SrFe_2As_2$ left $T_c^{onset}$ and $\rho_{ab}$ just above $T_c$ unchanged and sharpened the width of the transition in our Sample #4 by approximately a factor of two, as shown in Fig. 2. Based on the current sensitivity of the superconductivity, and the result on one sample discussed above in the Experimental section where peeling and recontacting the surface affected superconductivity (see inset to Fig. 2), perhaps some sort of planar (possessing the anisotropy of the crystal) superconductivity at or near the surface involving self-doping via defects is present. Such a mechanism would affect the bulk $\rho_{ab}$ values in the normal state only marginally, explaining the lack of correlation between the normal state residual resistivity $\rho_{ab}(T\rightarrow0)$ values and the occurrence of the observed 'partial' superconductivity. Preliminary measurements on a sample with current in the c-axis direction indicated no superconductivity. Further measurements to investigate the cause of superconductivity on larger samples from better optimized growth batches are underway.

**4. Conclusions**

Sample dependent superconductivity at $T_c \sim 20$ K with low critical current densities indicative of restricted dimension is observed in undoped In-flux grown single crystals of $BaFe_2As_2$. This superconductivity shows the same apparent anisotropy in its critical magnetic fields as bulk samples, and remains after the same annealing regimen that destroys superconductivity in undoped $SrFe_2As_2$.

Acknowledgements: The authors gratefully acknowledge helpful discussions with Art Hebard, Peter Hirschfeld, Pradeep Kumar, Dmitri Maslov, Johnpierre Paglione, Shanta Saha and Joe Thompson. Thanks as well to Johnpierre Paglione for supplying their numerical results for $H_{c2}(T)$ for $SrFe_2As_2$ shown in Fig. 3. Work at Florida performed under the auspices of the United States Department of Energy, contract no. DE-FG02-86ER45268.

Fig. 1 (Color online) Resistivity vs temperature for five samples of In-flux grown single crystals of $BaFe_2As_2$, with samples 1, 2, and 3 showing full superconducting transitions of $\rho$ to 0, while the resistivity of sample #4 approaches the finite value of 0.01 m$\Omega$-cm as T$\rightarrow$0 and sample #5 remains normal. Currents used were 0.1 mA except for sample #2 (1 mA) and sample #5 (1.5 mA).

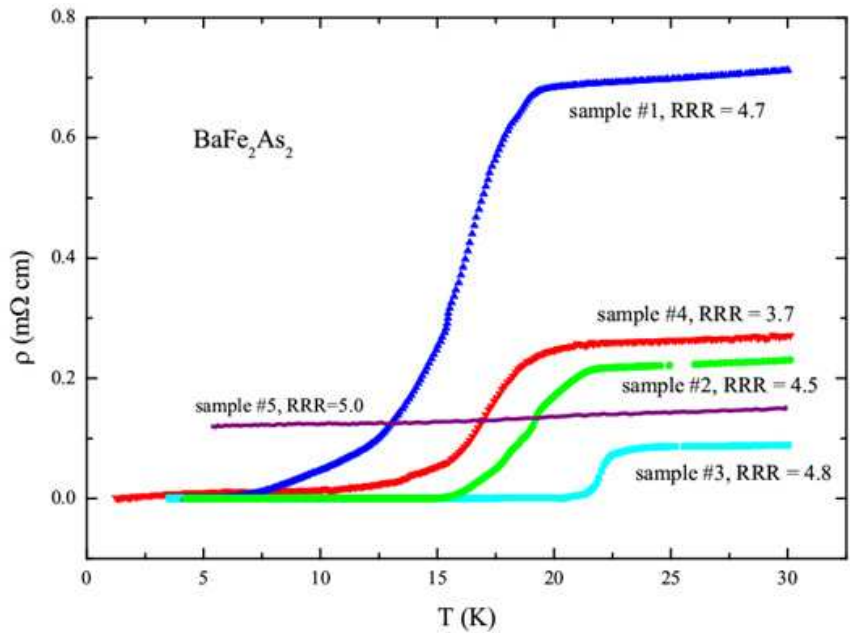

Fig. 2 (Color online) Resistivity vs temperature as a function of current for samples 1, 3 and 4. The solid symbols are for I=0.1 mA like shown in Fig. 1; for higher currents (1.5, 1.0, and 1.0 mA respectively), the resistive data (represented by solid lines) show significantly broader superconducting transitions. In sample 1, the higher current (corresponding to a current density of only 1.5 A/cm$^2$) actually prevents ρ from falling to 0 above 4 K. Annealing sample #4 (300 $^o$C for 2 hours under vacuum sealed in pyrex), solid red diamonds, sharpens the transition a factor of ~2 while changing neither $T_c^{onset}$ (which is very gradual in the unannealed sample) nor the measured finite value of ρ as T→0. The growth of the small feature around 14.5 K in the unannealed sample #4 into a clear shoulder almost 3 K broad centered at 18 K in the annealed sample is under investigation. The inset shows the resistivity vs temperature of sample #3 before (solid circles) and after (solid squares) peeling and recontacting. As discussed in the text, the fact that sample #3 shows a sharper, higher $T_c$ after peeling and being recontacted may imply a surface effect.

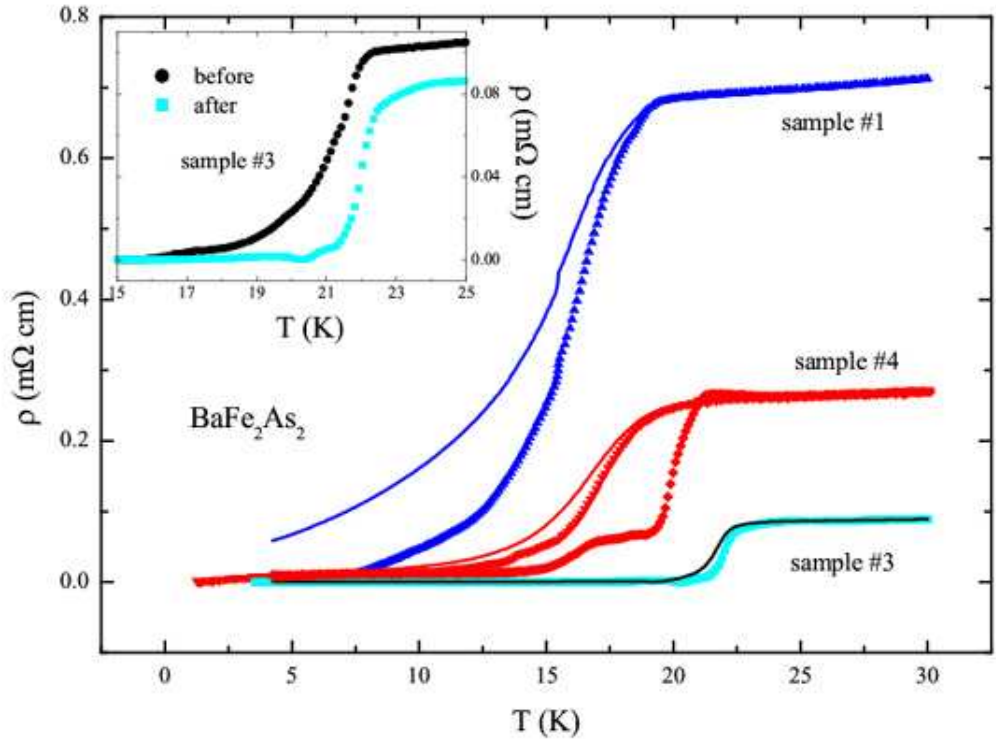

Fig. 3 (Color online) Critical field for sample #1, undoped $BaFe_2As_2$, for [8] undoped $SrFe_2As_2$, and for [15] $BaFe_{1.8}Co_{0.2}As_2$. Note the similar slopes for all three samples in both crystalline directions.

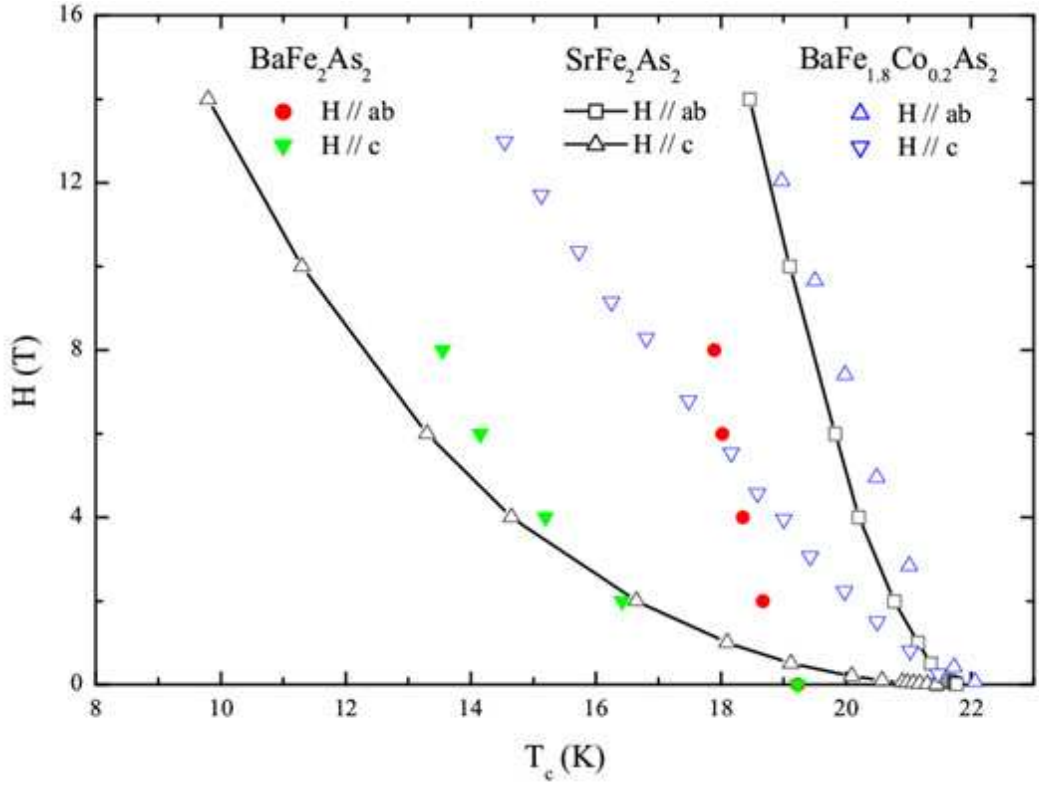

# References

[†]REU student, University of Michigan
*Corresponding Author